\documentclass[preprint,prx,longbibliography]{revtex4-1}
\usepackage{graphicx}
\usepackage{comment}
\usepackage{amsmath,amssymb,amsthm} 
\usepackage{bm}
\usepackage{verbatim}
\usepackage{float}

\begin{document}

		\title{ Newtons discrete dynamics}

\author{ S\o ren  Toxvaerd }
\affiliation{ Department
 of Science and Environment, Roskilde University, Postbox 260, DK-4000 Roskilde, Denmark}
\date{\today}

\vspace*{0.7cm}

\begin{abstract}
In 1687 Isaac Newton published   PHILOSOPHI\AE \ NATURALIS PRINCIPIA MATHEMATICA, where the classical analytic dynamics was
formulated. But Newton also formulated  a discrete dynamics, which is the central difference algorithm, 
known as the Verlet algorithm. 
In fact Newton used the central difference to derive his second law.

	The central difference algorithm is used in computer simulations,
	where almost all Molecular Dynamics simulations are performed with the Verlet algorithm or other reformulations of the central difference algorithm.
 Here we show, that the discrete dynamics obtained by Newtons algorithm for Kepler's equation 
has the same solutions as the analytic dynamics. The discrete positions of a  celestial body  are located on an ellipse, which is
the exact  solution for  a  shadow Hamiltonian nearby the Hamiltonian for the analytic solution. 

\end{abstract}

\maketitle

\section{Introduction}

In 1687 Isaac Newton published PHILOSOPHI\AE \ NATURALIS PRINCIPIA MATHEMATICA.  $(Principia)$ \cite{Newton1687}
with the foundation of  the classical  analytic dynamics.
Newton described the dynamics of an  object by means of a differential equation,
 and in the Lagrange-Hamilton formulation of the classical dynamics the
 position $\textbf{r}(t)$ and momentum  $\textbf{p}(t)$ are analytic dynamical variables of a coherent   time. 
 But  in $Principia$ Newton also derived  a discrete dynamics, where a celestial body's 
 positions are obtained at discrete times. The discrete Newtonian dynamics has the same
 invariances as the analytic dynamics, but differs fundamentally by that only the discrete positions
 are dynamic variables of the discrete time.

Today almost all numerical integrations of classical dynamics are performed by Newtons discrete dynamics, by updating the positions
at discrete times. The  Newtonian  dynamics is the
classical limit dynamics of  the relativistic quantum dynamics, and 
 the fundamental length and time  in  quantum dynamics are
the  Planck length $l_{\textrm{P}} \approx 1.6 \times 10^{-35}$m and 
 Planck time $t_{\textrm{P}} \approx 5.4 \times 10^{-44}$ s \cite{Garay}.
They are immensely smaller than the differences in step lengths
 and the time increments used in  the numerical integration by discrete dynamics,  so the difference 
 between the two dynamics in the classical limit for the dynamics of heavy objects
 with slow motions is $ nihil$.

Newtons 
 discrete dynamics  has the same qualitative behaviour as the analytic. It is time
reversible, symplectic \cite{Toxvaerd1993}, and has the same invariances as the analytic dynamics: conservation of  momentum, angular momentum and energy \cite{Toxvaerd2014}. It is furthermore possible by an asymptotic expansion to make it probably,
that the  positions of an object obtained by Newtons
discrete dynamics are located on the analytic  trajectory  for a $shadow$ $Hamiltonian$ nearby the Hamiltonian for
the corresponding analytic dynamics \cite{toxone}. If that is the case the numerical
generation of positions in computer simulations (Molecular Dynamics) is  the exact positions for the discrete dynamics
obtained by Newtons central difference algorithm.
Here we show that the dynamics, obtained by solving Keplers equation  for celestial objects by  discrete dynamics, 
give stable orbits which only differ marginally from the corresponding analytic orbits and with a strong indication of
a  shadow Hamiltonian for the dynamics.

\section{Newtons discrete dynamics: The central difference algorithm}

Newtons  second law  relate an object with mass \textit{m}  at the position, \textbf{r}(t), momentum, \textbf{p}(t),
 at time t 
with the force  \textbf{F(r)}. The English translation \cite{Newtonengtrans} of the Latin formulation of Newtons second law second law is

\textit{ The alteration of motion}(momentum) \textit{is
ever proportional to the motive force impressed; and is made in the direction of the right line in which that force
is impressed.}, i.e.

\begin{equation}
	\bf{F}(r)=\frac{d\bf{p}}{\it{dt}}, 
\end{equation}
and in \textit{Section II}, Newton derived an interesting relation:

\textit{PROPOSITION I. THEOREM I.}
\textit{The areas, which resolving bodies describe by radii drawn to an immovable centre of
force do lie in the same immovable planes, and are proportional to the times in which they are described.}

 Newton noticed, that (see Figure 1):
\textit{For suppose the time to be divided into equal parts, and in the first part of that time
let the body by its innate force describe the right line AB. In the second part of that time,
the same would (by Law I.), if not
hindered, proceed directly to c, along the line Bc equal to AB; so that by
the radii AS, BS, cS, drawn to the centre, equal areas ASB, BSc, would be described.
But when the body is arrived at B, suppose that a centripetal force acts at once with a great impulse, and, turning aside the body
from the right line Bc, compels it afterwards to continue its motion along the right line BC. Draw cC parallel to BS
meeting BC in C;
and at the end of the second part of the time, the body (by Cor. I. of the Laws) will be found in C,
in the same plane with triangle ASB Join SC, and, because SB and Cc are parallel, the triangle SBC will
be equal to the triangle SBc, and therefore also to the triangle SAB.}

  \begin{figure}
 	 \includegraphics[width=8.6cm,angle=0]{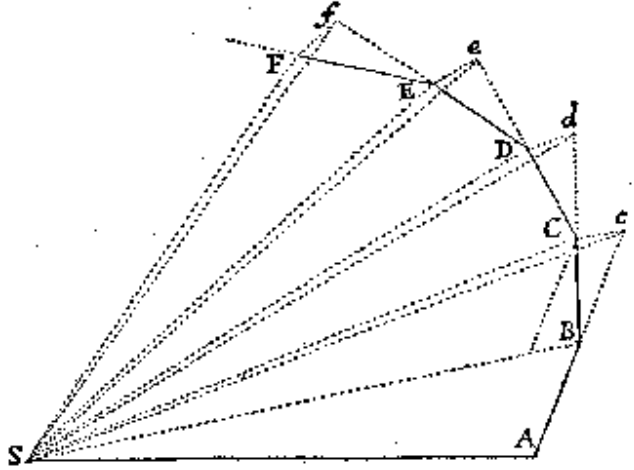}
 	 \caption{  Newton's figure in Principia, at his formulations of the discrete dynamics. 
	  The discrete positions are A: $\bf{r}_A$$(t_0)$;  B: $\textbf{r}_B(t_0+\delta t)$;
	  C: $\textbf{r}_C(t_0+2\delta t)$, etc.. The deviation 
	  from the straight line ABc (Newton's first law) to point C is caused by a  force at S, acting on the object at  
	  point B at time $t_0+ \delta t$.}
  \end{figure}

 So according to Newton's $PROPORSITION$ the  particle moves with constant momentum, 
$m(\textbf{r}_B(t_0+\delta t)- \textbf{r}_A(t_0))/\delta t$  from the position  $\textbf{r}_A(t_0)$ to the position   $\textbf{r}_B(t_0+ \delta t)$ in the
time interval $t \in [t_0,t_0+\delta t]$, where a force, $\textbf{F}(\textbf{r}_B)$  instantaneously changes the momentum.
This formulation of the discrete updating of positions: $\textbf{r}_A(t_0)$, $\textbf{r}_B(t_0+\delta t)$ $\textbf{r}_C(t_0+2\delta t)$,..
with constant momentum in the time intervals between the updating is \textit{the central difference algorithm}
\begin{equation}
	m\frac{\textbf{r}(t_n+\delta t)-\textbf{r}(t_n)}{\delta t}
	=m\frac{\textbf{r}(t_n)-\textbf{r}(t_n-\delta t)}{\delta t} +\delta t \textbf{F}(t_n).	
\end{equation}
The algorithm determines the $n+1$'the position from the two previous positions by
\begin{equation}
	\textbf{r}(t_n+\delta t)=2\textbf{r}(t_n)-\textbf{r}(t_n-\delta t) +\frac{\delta t^2}{m} \textbf{F}(t_n).
\end{equation}
and this formulation of Newton's central difference algorithm is  the so called "Verlet" algorithm \cite{Verlet1967,Levesque2018}, which is
  used in Molecular Dynamics simulations \cite{Hockney,AllenandTildesley,FrenkelandSmit}.
The algorithm can be reformulated, if one updates the positions in two steps   with
$	\textbf{v}(t_n+\delta t/2) \equiv (\textbf{r}(t_n+\delta t)-\textbf{r}(t_n))/\delta t$:
	\begin{eqnarray}
		\textbf{v}(t_n+\delta t/2) =\textbf{v}(t_n-\delta t/2)+\frac{\delta t}{m} \textbf{F}(t_n) \nonumber \\
		\textbf{r}(t_n+\delta t)=\textbf{r}(t_n)+ \delta t 	\textbf{v}(t_n+\delta t/2), 
	\end{eqnarray}		
and this reformulation is named the  "leap-frog" algorithm.  It is the discrete version of Euler's equations for Newtons
analytic dynamics \cite{Cromer1981}.

There are several things to note about   Newtons formulation of the discrete dynamics. According to Newton the
 force acts \textit{  at once with a great impulse}, i.e. the forces are discrete, it acts only at the
 discrete times $t_n$ and the object is not expose to the force within the time intervals between the discrete times where
 it moves with constant momentum  as Newton explicit notes: \textit{(by Law I.)}.

 Another thing to note is, that Newtons in $Principia$ did not write that the
 constant areal of the triangles is Keplers second law for the  planets orbits around the Sun. But Newton must have
 notices  this fact and must have realized that his dynamics, even in the discrete version,
 most likely could explain the celestial dynamics. The equal area of the triangles and Keplers second law
 is according to the proof in \textit{PROPOSITION I. THEOREM I.}
 valid for any central force   between two celestial objects. 
It is a consequence of the conserved angular momentum in the discrete and analytic dynamics
 (see later).
 The  $Principia$  is written long time after Newton in fact had formulated his classical dynamics, and Newton solved
 Keplers equation (geometrically!)  for the analytic dynamics in  $Principia$.

 A third thing to note is the continuation  of \textit{PROPOSITION I. THEOREM I.}:
    \textit{Now let the number of those triangles be augmented, and their breadth diminished 
    in infinitum; and (by Cor.4, Lem, III) their ultimate perimeter ADF will be a curve line:
    and therefore the centripetal force, by which the body is perpetually drawn back
    from the tangent of this curve, will act continually; and any
    described areas SADS, SAFS, which are always proportional to the times
    of description, will, in this case also, be proportional to those times. Q.E.D.}.
    So Newton used the central difference to obtain his analytic dynamics and
    noticed, that by letting the time increment go to zero he obtained not only a
    curve line and  a continuous force, but also maintained the constant area of the triangles.
    But he did not mentioned Keplers second law.

There exists several  other reformulations of the central difference algorithm \cite{ AllenandTildesley,FrenkelandSmit}. 
The Verlet algorithm was derived by L. Verlet by a forward and backward Taylor expansion, and the algorithm and its
many reformulations are normally presented as a third order predictor of the positions, obtained
by Taylor expansions.  Newton was well aware of Taylor expansions; but he did not used it
to formulate a discrete dynamics. It is the other way around, Newton used the discrete dynamics
to obtain the analytic dynamics and his second law.

Before the formulation of the discrete dynamics for a celestial body is presented, the solution
of Keplers equation for analytic dynamics is summarized in the next section.  

\section{The solution of Kepler's equation}
\subsection{The analytic solution of Kepler's equation}
Newton solved in $Principia$, Kepler's equation for the orbit of a planet. 
The solution of Kepler's equation \cite{Tokis2014}
\begin{equation}
	\frac{d^2\textbf{r}(t)}{dt^2}=-\frac{g Mm}{r(t)^2}\hat{\textbf{r}}
\end{equation}
for a planet with the gravitational constant $g$ and mass $m$ at the position $\textbf{r}(t)$ from the  Sun at the origin \cite{explain} with mass \textit{M} 
relates
the constant energy, 
\begin{equation}
	E=1/2 m\textbf{v}(t) \cdot \textbf{v}(t) -gMm/r(t),
\end{equation}
with the semi major  axis in an ellipse

\begin{equation}
	a=-gMm/2E.
\end{equation}
The longest distance $r_{max}$ (aphelion) from the Sun is
\begin{equation}
	r_{max}=2a-r_p,
\end{equation}
where $r_p$ is the shortest distance (perihelion)  to the Sun.
The  eccentricity, $\epsilon$, is
\begin{equation}
	\epsilon=\frac{r_{max}-r_p}{r_{max}+r_p}=1-\frac{r_p}{a}.
\end{equation}
and the semi minor axis, $b$ is
\begin{equation}
b  =a\sqrt{1-\epsilon^2}.
\end{equation}
With the major axis in the $x$-direction the planet moves in a stable elliptic orbit
\begin{equation}	
	\frac{(x(t)-(a-r_p))^2}{a^2}+\frac{y(t)^2}{b^2}=1,
\end{equation}
for
\begin{equation}
0 \leq \epsilon < 1,
\end{equation}
within a orbit period
\begin{equation}
	T(orbit)=2\pi\sqrt{a^3/gM}.
\end{equation}	
The velocity  at perihelion, $\textbf{v}_p(t))=(0,vy_p$),  is in the $y$-direction and the energy is 
\begin{equation}
	E=1/2mvy_p^2-gMm/r_p, 
\end{equation}
and since $1/a=-2E/gMm=-mvy_p(t)^2+2/r_p$, the limit values for elliptic orbits can be expressed by the maximum velocity as
\begin{equation}
	\sqrt{gM/r_p} \leq vy_p < \sqrt{2gM/r_p}.
\end{equation}

Let the planet at time $t_0=0$ be in the perihelion  of the elliptic orbit with the maximum velocity  $\textbf{v}_{p}=(0,vy_p)$ at the shortest
distance, $\textbf{r}_{min}=(x(t_0),y(t_0))=(-r_p,0)$, from the Sun, which is located at the origin.
  The classical orbit of a  planet can be obtained from these four parameter: $ gM, m, r_p, vy_p  $ ( or:  $ gM, m, r_{max}, vy_{min}  $ at aphelion).

\subsection{Kepler's orbit obtained by Newton's  central difference algorithm }

The discrete dynamics can be obtained from the same parameters, $ gM, m, r_p, vy_p $
together with the discrete time increment $\delta t$.
Newton's   discrete dynamics for the $n+1$'th change of position of a planet  is
\begin{equation}
	\frac{\textbf{r}(t_{n+1})-\textbf{r}(t_n)}{\delta t}=\frac{\textbf{r}(t_{n})-\textbf{r}(t_{n-1})}{\delta t}
	-\frac{gMm \delta t}{r(t_n)^2}\hat{\textbf{r}}(t_n)
\end{equation}
An important fact is, that the algorithm relates a new position with the two
previous positions and the forces at the time, where the forces act. I.e., the momentum (or velocity) is not a
dynamical variable in the discrete dynamics, and any expression
for velocity, and thereby the kinetic energy  is ad hoc.

The discrete time evolution with the constant time increment $\delta t$, obtained by Newton's central difference algorithm,
starts from either two sets of  positions,
$\textbf{r}(t_0),\textbf{r}(t_0-\delta t)$ (Verlet algorithm),
or, as Newton illustrated, from a position $\textbf{r}(t_0)$ and a previous
change of position $ \textbf{r}(t_0)-\textbf{r}(t_0-\delta t) \equiv \delta t \textbf{v}(t_0-\delta t/2)$,
in the time interval $t \in [t_0-\delta t,t_0]$ (Leap frog or implicit Euler algorithm).
The velocity $\textbf{v}(t_n)$ at  the time where the force acts, at  the position   $\textbf{r}(t_n)$, is   in general obtained by a central difference
\begin{equation}
	\textbf{v}(t_n)= \frac{\textbf{v}(t_n+\delta t/2)+\textbf{v}(t_n-\delta t/2)}{2}=
	\frac{\textbf{r}(t_n+\delta t)-\textbf{r}(t_n-\delta t)}{2\delta t}.
\end{equation}

Newton's discrete time reversible dynamics has the same three invariances as his analytic dynamics.
It conserves the (total)  angular momentum, $\textbf{L}$.
 The angular momentum, $\textbf{L}(t_n)$ for  a planet at
the $n$'th time step (and using the Verlet-formulation, Eq. (17) and the fact, that the force is in the direction of the discrete position) is
\begin{eqnarray} 
	\frac{2 \delta t}{ m} \textbf{L}(t_n)=\textbf{r}(t_n) \times (\textbf{r}(t_{n+1}) -\textbf{r}(t_{n-1}) )  \nonumber \\
=\textbf{r}(t_n) \times (2\textbf{r}(t_{n}) -2\textbf{r}(t_{n-1}) )  \nonumber \\
	=\textbf{r}(t_n-1) \times (\textbf{r}(t_{n}) +\textbf{r}(t_{n})) =  \nonumber \\
	=\textbf{r}(t_n-1) \times (\textbf{r}(t_{n}) -\textbf{r}(t_{n-2}))=	\frac{2 \delta t}{ m}\textbf{L}(t_{n-1}).
\end{eqnarray}	
It is straight forward to prove, that the  
 constant area of the triangles in Newton's formulation of  the discrete dynamics (Figure 1) 
 is a consequence of the conserved angular momentum.

If one determines the  energy at the $n$'th time step by
\begin{equation}
	E_{disc}(t_n)= \frac{1}{2}m\textbf{v}(t_n)^2-\frac{gMm}{r(t_n)},
\end{equation}
it fluctuates during the discrete time  propagation, although the mean value
remains constant.

\subsection{The shadow Hamiltonian for the central difference algorithm }

The points obtained by Newton's central difference algorithm for a simple harmonic force 
is located  on a harmonic trajectory of a  harmonic "shadow Hamiltonian" $\tilde{H}(\textbf{q,p})$  \cite{toxone},
 with position $\textbf{q}$ and momentum $\textbf{p}$ in the Lagrange-Hamilton equations.
The shadow Hamiltonian $\tilde{H}$ for a symplectic and time-reversible  discrete algorithm
 can in general be obtained from the corresponding $H(\textbf{q,p})$ for the  analytic dynamics by an asymptotic
expansion in the time  increment $\delta t$, if the potential energy is analytic \cite{Sanz-Serna,Hairer,Reich},

\begin{equation}
\tilde{H} = H+ \frac{\delta t^2}{2!}g(\textbf{q},\textbf{p})+ \mathcal{O}(\delta t^4),
\end{equation}
The corresponding  energy invariance, $\tilde{E}$, for the discrete dynamics
 in Cartesian coordinates for $N$  particles is \cite{toxone,Gans,toxtwo}
\begin{equation}
	\tilde{E_n} 
	= U(\textbf{R}_n) + \frac{1}{2} m \textbf{V}_n^2 +
 \frac{\delta t^2}{12} \textbf{V}_n^T
 \textbf{J(R}_n) \textbf{V}_n - \frac{\delta t^2}{24m}\textbf{F}_n(\textbf{R}_n)^2  \\
	+ \mathcal{O}(\delta t^4),
\end{equation}
 where  $\textbf{J}$ is the Hessian, $\partial^2  U(\textbf{q})/\partial \textbf{q}^2$,
  of the potential energy function $U(\textbf{q})$, the velocity
 of the $N$ particles is 
 $\textbf{V}_n\equiv(\textbf{v}_1, ..., \textbf{v}_N)$, and the force 
 with position  $\textbf{R}\equiv(\textbf{r}_1, ..., \textbf{r}_N)$ is
 $\textbf{F}(\textbf{R})\equiv(\textbf{f}_1(\textbf{R}), ..., \textbf{f}_N(\textbf{R}))$.

The observed  energy fluctuations  for a complex system decreases by a factor of  hundred or even more
by including these terms in the expression for the  energy
and it indicates, that the  expansion is rapidly converging for relevant time increments \cite{toxtwo,toxa}.

 The shadow energy at the n'th  step  for a planet, attracted by the Sun at a fixed position at the origin, can be obtained from the
 expressions in Appendix A in \cite{toxtwo}. It is
\begin{equation}
	\tilde{E}(t_n)=E_{disc}(t_n) 
	-\frac{\delta t^2}{12} \left
	( \frac{3gMm}{r(t_n)^5}(\textbf{v}(t_n)\textbf{r}(t_n))^2+\frac{gMm}{r(t_n)^3}\textbf{v}(t_n)^2 \right ) 
  -\frac{\delta t^2(gMm)^2}{24 r(t_n)^4} 	+ \mathcal{O}(\delta t^4).
\end{equation}

 \section{The  orbit of a planet obtained by Newton's discrete algorithm }

The  positions of a planet are obtained by Newton's central difference algorithm. The  positions are determined by the time increment $\delta t$
and by the same parameters as the analytic curve,
e.g. $gM, m, r_p$ and $vy_p$.  The  curves through  the  points are almost identical to the analytic ellipses, and  the discrete dynamics
obeys the same condition  for a stable elliptic orbit  as the analytic dynamics (Eq. (15)). Figure 2 shows the orbits, obtained with different start values of the velocity, $vy_p$ \cite{units}.

  \begin{figure}
 	 \includegraphics[width=6cm,angle=-90]{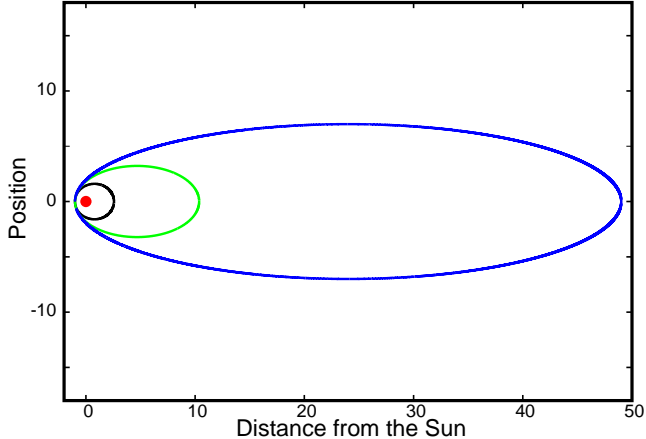}
 	 \caption{ The orbits of an Earth-like planet. The curves are obtained by Newton's central
	  difference algorithm  from the perihelion at $\textbf{r}(t_0)=(-1,0)$ with
	  $gM=m=1$, and with different velocities $vy_p$. The red filled circle is the position of the Sun
	  and the three curves are for $vy_p$= 1.2 (black); 1.3 (green) and 1.4(blue), respectively.}
  \end{figure}	

  The generation of  positions by the central difference algorithm needs either
  two consecutive start positions, $\textbf{r}(t_0-\delta t)$ and $\textbf{r}(t_0)$,
  or $\textbf{r}(t_0)$ and $\textbf{v}(t_0-\delta t/2)$.
  It is convenient to start the dynamics in perihelion (or aphelion) where $vx(t_0)=0$.
Due to the  time reversibility of the discrete dynamics 
$vy(t_0+\delta t/2)=vy(t_0-\delta t/2)$ and   $vx(t_0+\delta t/2)=-vx(t_0-\delta t/2)$ at perihelion. The first discrete position
away from the perihelion, $x(t_0+\delta t),y(t_0+\delta t)$, is
\begin{equation}
	\textbf{r}(t_0+\delta t)=	x(t_0+\delta t),y(t_0+\delta t)= -r_p+\frac{1}{2}\frac{gM \delta t^2}{r_p^2},\delta t vy_p,	
\end{equation}
and since $x(t_0+\delta t)=x(t_0-\delta t)$ due to the time symmetry,  the discrete dynamics starts with an energy  $ E_{disc}(t_0)$ at   time $t_0$=0,   which is equal to the constant
energy $E$ in the analytic dynamics.

\subsection{ A shadow Hamiltonian and  the functional form of the orbits for the discrete dynamics}

The question is: Is there a shadow Hamiltonian for the discrete dynamics of a planet's orbital motion,
and if so, what is the functional form of the  analytic function for $\tilde{H}$. Since  the 
discrete dynamics for $\delta t$ going to zero converges to the analytic dynamics with elliptic motion, it is natural to fit an ellipse to the discrete points.

The main investigation is for an Earth-like planet with $vy_p=1.2$ at $r_p=-1$ and with $gM=m=1$.
The results are given in Table I. with data for different values  of the number $n$ used to
integrate one orbit, $n = T(orbit)/ \delta t$, where   $T(orbit)$  is the
orbit time with analytic dynamics (Eq. (13)).
The investigation shows several things.

The discrete points are with high precision on an ellipse even for relative few number of integration points $n$. Figure 3 shows the planet's positions near perihelion and
when the position is updated every $\delta t= T(orbit)/365$,  or $\approx$ 24 hours for
an Earth-like planet. Column 2 and 3 in the Table give the fitted values for the axes and with the
rms stand deviations of the fits in column 4. E. g. a deviation of $3. \times 10^{-8}$ corresponds to $\approx$ 3-4 km in the case of planet Earth.

  \begin{figure}
 	 \includegraphics[width=6cm,angle=-90]{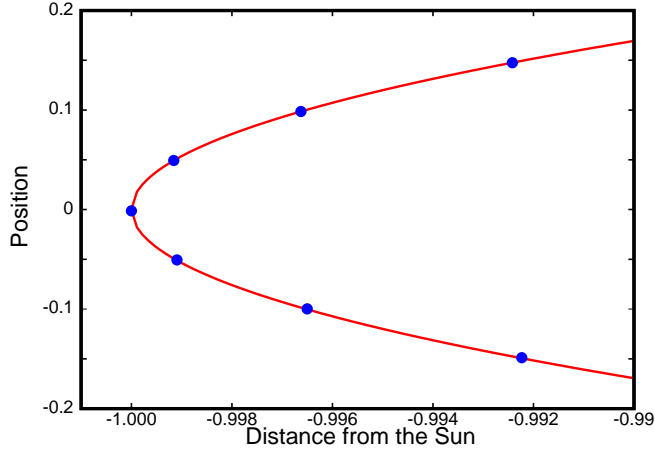}
 	 \caption{ The discrete positions of an Earth-like planet near perihelion. The discrete positions (red filled circles) are obtained by Newton's central
	  difference algorithm  with $\delta t=T(orbit)/365$, i.e. for an Earth-like planet every 24 hours. The full line is an ellipse determined
	  from the 365 discrete points by fitting the axes of an ellipse.}
  \end{figure}	

The  mean energies, $<E_{disc}>$ and  $<\tilde{E}_{disc}>$ are given in column 5 and 6.
The observed energy fluctuations are decreased by a factor of the order $\approx 10^3$ to $10^5$ just by inclusion of the first order correction 
(Eq. (22)). Figure 4 shows the energy evolution during tree times in the orbit. The tiny energy variations of the shadow energies are shown in the insert.

The discrete dynamics was obtained for other values of $vy_p, gM. m$ and  $r_p$ and confirmed the result, that the discrete dynamics
behaves as the analytic. The discrete positions were located on  ellipses and the energies, $\tilde{E}(t_n)$  were almost constant by inclusion of
the first order term (Eq. (22)) in $E_{disc}(t_n)$.


 \begin{center}
	 \textbf{Table 1}. Principal axis and discrete energies for  $r_p=-1$, $vy_p=1.2$, $gMm/r_p=-1$ and $\delta t=T(orbit)/n$. 
 \end{center}
\begin{tabbing}
\hspace{1.cm}\=\hspace{3.cm}\=\hspace{3cm}\=\hspace{2cm}\=\hspace{5cm}\= \hspace{2cm} \\
 $n$   \> Major axis  \> Minor axis\  \> rms   \> $E_{disc}$  \> $\tilde{E}$  \\
---------------------------------------------------------------------------------------------------------------------------\\
365     \>  1.7867062  \> 1.60399     \> 4.$\times 10^{-4}$ \> -0.27988$\pm 3.10^{-5}$  \>-0.2798678  $\pm 1.10^{-8}$     \\
$10^3$  \>  1.7858364 \> 1.603624    \> 2. $\times 10^{-4}$ \> -0.279984  $\pm 3.10^{-6}$  \> -0.2799823897   $\pm 3.^{-10}$ \\
$10^4$  \>  1.7857156  \> 1.603568016 \> 2. $\times 10^{-7}$ \>  -0.27999984   $\pm 4.10^{-8}$  \> -0.2799998239055  $\pm 7.10^{-13}$ \\
 $10^5$ \>  1.78571423 \> 1.603567457 \> 3. $\times 10^{-8}$ \>  -0.2799999985   $\pm 4.10^{-10}$  \>-0.27999999823913   $\pm 1.10^{-14}$ \\
$\infty$ \> 1.78571429 \> 1.603567451 \> 0 \>  -0.28 \>-0.28 \\
---------------------------------------------------------------------------------------------------------------------------\\
\end{tabbing}

 \begin{figure}
 	 \includegraphics[width=6cm,angle=-90]{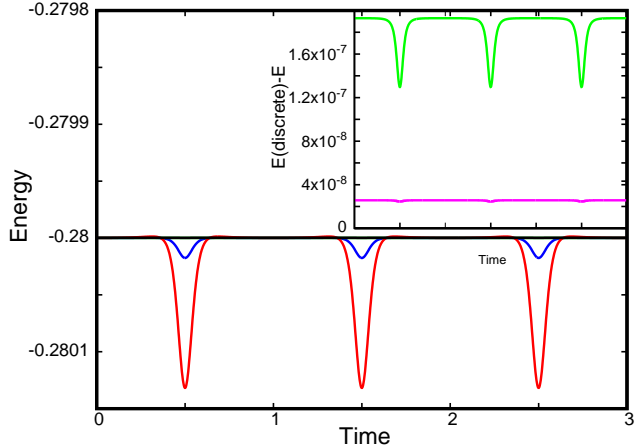}
	  \caption{ The  energies $E(t_n)$ and $\tilde{E}(t_n)$ for for the circulation of a planet tree times in its elliptic orbit.
	  The discrete values are obtained by starting from the aphelion $\textbf{r}(t_0)=(r_{max},0)$ with $r_{max}$ and $vy_{min}$ obtained from
	  $r_p=-1$, $vy_p=1.2$, $gM=m=1$ and
	  $\delta t_1=T(orbit)/365$ and  $\delta t_2=T(orbit)/1000$ ($\approx$ one day and eight hours, respectively).
	  Red: $E(t_n)$ with  $\delta t_1$; blue: $E(t_n)$ with  $\delta t_2$; black: $E(analytic)=-0.28$. The inset
	  shows the small energy differences between the corresponding shadow energies, $\Delta E(t_n)=\tilde{E}(t_n)-E(analytic)$.
	  Green:  $\Delta E(t_n)$ for $\delta t_1$; Magenta:   $\Delta E(t_n)$ for $\delta t_2$.  }
  \end{figure}

 \section{Discussion}

 The Molecular Dynamics simulations  strongly indicate, that there exists a
 shadow Hamiltonian for the discrete Newtonian dynamics of celestial bodies.
 The existence of a shadow Hamiltonian for the discrete dynamics   implies that the positions,
 obtained by Newtons  discrete dynamics
  are exact and with the same dynamics invariances as the analytic
  dynamics: conservation of momenta, angular momenta and total energy. 
 But despite the same dynamic invariances, there
is, however, one fundamental difference between the two dynamics.
Only the  positions and time are variables in the discrete dynamics, the momenta are not.

  Newton used the central difference algorithm to derive his second law for classical dynamics,
 but he newer, in $Principia$, calculated  a celestial body's positions by using the  algorithm.
Isaac Newton and Robert Hooke used, however, the geometric implementation (Figure 1) of the central difference
algorithm to construct a celestial body's orbit \cite{Nauenberg2018};
but they were of course not aware of, that the discrete dynamics
has the same qualities as Newton's analytic dynamics.

 The Newtonian analytic dynamics have been questioned.
  T. D. Lee and coworkers have  
 analysed discrete dynamics in a series of publications.
 The analysis covers not only classical mechanics \cite{Lee1}, but also non relativistic quantum mechanics
 and relativistic quantum field theory  \cite{Lee3}, and  Gauge theory and Lattice Gravity \cite{Lee2}.
The discrete dynamics is obtained by treating positions and time, but not momenta, as a discrete dynamical variables
as in Newtons discrete dynamics.
The Newtonian dynamics has also been modified ad hoc by M. Milgrom \cite{mond} in order to explain the stability of
galaxies.

 The  indication of the exactness of Newtons  discrete dynamics  raises the principle question:
 Which of these two formulations  is  the  correct classical limit law for relativistic quantum dynamics?
  The momenta and positions in the discrete dynamics are asynchronous as is the case in quantum dynamics, but
the difference, in the classical limit  between the two formulations is, however, immensely small.
If the  discrete dynamics  is the  correct formulation, Newton will also be the founder of this dynamics.

				\section{Acknowledgment}
 Ole J. Heilmann, Niccol\~{o} Guicciardini and Jeppe C Dyre is gratefully acknowledged.
This work was supported by the VILLUM Foundation’s Matter project, grant No. 16515.


\begin{thebibliography}{99}
\bibitem{units}  It is convinient to express the energy and length in units
 of a given planet energy, e.g. $E^*=gMm^*/r_p^*$,  and length $r_p^*$ from the sun (e. g. the planet Earth).
 The corresponding time unit is $t^*=r_p^*\sqrt{m^*/E^*}$. The relations below are given in these  reduced units.
\end{thebibliography}
\end{document}